# Extending the spectral operation of multimode and polarization-independent power splitters through subwavelength nanotechnology

Raquel Fernández de Cabo[a,*], David González-Andrade[b], Pavel Cheben[c], Aitor V. Velasco[a],

[a] *Instituto de Óptica Daza de Valdés, Consejo Superior de Investigaciones Científicas (CSIC), 121 Serrano, Madrid 28006, Spain*
[b] *Centre de Nanosciences et de Nanotechnologies, CNRS, Université Paris-Saclay, Palaiseau 91120, France*
[c] *National Research Council Canada, 1200 Montreal Road, Bldg. M50, Ottawa K1A 0R6, Canada*
* Corresponding author: r.fernandez@csic.es

**ABSTRACT**

Power splitters play a crucial role in virtually all photonic circuits, enabling precise control of on-chip signal distribution. However, state-of-the-art solutions typically present trade-offs in terms of loss, bandwidth, and fabrication robustness, especially when targeting multimode operation. Here, we present a novel multimode 3-dB power splitter based on a symmetric Y-junction assisted by two regions of subwavelength grating metamaterials with different geometries. The proposed device demonstrates high-performance for multimode and dual-polarization operation with relaxed fabrication tolerances by leveraging the additional degrees of freedom offered by two distinct geometries of subwavelength metamaterials to control mode evolution. Our design achieves, to the best of our knowledge, the widest operational bandwidth reported to date for a nanophotonic multimode silicon power splitter. Simulations for a standard 220-nm-thick silicon-on-insulator platform predict minimal excess loss (< 0.2 dB) for the fundamental and the first-order transverse-electric modes over an ultra-broad 700 nm bandwidth (1300 – 2000 nm). For the fundamental transverse-magnetic mode, losses are less than 0.3 dB in the 1300 – 1800 nm range. Experimental measurements validate these predictions in the 1430 – 1630 nm wavelength range, demonstrating losses < 0.4 dB for all three modes, even in the presence of fabrication deviations of up to ±10 nm. We believe that this device is suitable for the implementation of advanced photonic applications requiring high-performance distribution of optical signals, such as programmable photonics, multi-target spectroscopy and quantum key distribution.

## 1. Introduction

Silicon photonics has become a leading technology for the development of integrated optoelectronic circuits, benefiting from the well-established complementary metal-oxide-semiconductor ecosystem of the microelectronics industry [1]. This compatibility enables the cost-effective and large-scale fabrication of high-density photonic circuits. Such properties have boosted ultra-fast data transmission in integrated optical transceivers [2], and are now playing a crucial role in the development of a wide spectrum of emerging applications such as neuromorphic photonics, 5G communications, quantum photonics, the Internet of Things, light detection and ranging, spectrometry, and sensing [3–8]. However, addressing the needs of these applications requires the development of breakthrough fundamental silicon-integrated components. Amongst these components, power splitters provide one of the most essential and widespread functions, as they enable precise control of on-chip optical signal distribution and routing. To further illustrate, they play crucial roles in cutting-edge applications such as mode-division multiplexing, quantum key distribution, 6G communication networks, and neural networks [9–12]. Power splitters capable of operating over ultra-broad bandwidths with minimal losses are essential for high-capacity data transmission across multiple optical bands, multi-target spectroscopy, and programmable photonics, among other applications [13–16]. Specifically, these splitters are crucial in programmable photonics, where fixed hardware components require low losses and broadband optical responses with relaxed fabrication tolerances [15]. For high-capacity data transmission systems, power splitters must not only provide broad optical responses but also support higher-order optical modes for mode-division multiplexing [14] and dual polarization operation for polarization-division multiplexing systems [16]. Thus, the advancement of silicon photonic systems requires the development of compact 3-dB optical power splitters that offer low loss, broad bandwidth response, and operate for multiple modes and polarizations, while maintaining robust fabrication tolerances.

Many power splitting architectures have been proposed from the early days of integrated photonics to the present day, such as multimode interference (MMI) couplers, directional couplers, adiabatic devices, or Y-junctions. Focusing on progress made in recent years, various strategies have been reported to enhance MMI robustness against fabrication imperfections and mitigate the inherent wavelength dependence of these structures, including careful design of the multimode region to control the evolution of the optical field and waveguide core thickness increase [17]. In [18] an ultra-broadband and fabrication-tolerant splitter based on an MMI structure was proposed, but it was limited to monomode and single-polarization operation. Similarly, to cope with the narrow bandwidth of conventional directional couplers, shallow-etched rib waveguides and geometric optimizations (e.g., bent directional couplers or asymmetric directional couplers) have been employed [19]. These approaches provide in some cases polarization-insensitive operation with operational bandwidths restricted to 100 nm. Recent refinements in taper profiles of adiabatic devices have led to the demonstration of compact power splitters with lengths between 5 and 14 μm [20,21]. Despite these advancements, the experimental bandwidths achieved are still below 100 nm. Finally, several strategies have been proposed to circumvent symmetric Y-junctions performance degradation caused at the junction tip by the minimum feature size (MFS) limitations of current fabrication processes [22]. These solutions range from tapered and slotted waveguides [23–26], to particle swarm optimization algorithm [22,27], photonic crystals [28], or subwavelength gratings (SWG) [29–31].

Notably, since their first demonstration in silicon photonics, SWG metamaterials have turned into a powerful tool to precisely control light at the nanoscale [32]. These metamaterials consist of a periodic lattice of different dielectric materials spaced at a distance shorter than half the effective wavelength of the propagating light, which allows tailoring refractive index, dispersion, and anisotropy by modifying the geometrical parameters [33]. In addition to the demonstration of numerous devices with unprecedented performance [10,34–38], SWGs have also been successfully used to engineer different power splitters to achieve state-of-the-art performance [18,39–45]. As an illustrative example, ultra-broadband SWG MMIs for single-mode and single-polarization operation [18], compact polarization-independent three-guide SWG directional couplers [37], and high-performance SWG multimode Y-junctions [46] have been demonstrated. Other alternatives include inverse design-optimized pixelated meta-structures, showing promising prospects for broadband and compact devices although they have not yet been experimentally validated [47–49].

Despite the large number of power splitter architectures that have been proposed, none of these solutions offer compact designs that ensure low losses for high-order modes and both polarizations while operating across ultra-broadband wavelength ranges with relaxed fabrication tolerances. Here, we explore a new strategy to realize 3-dB power splitters in SOI satisfying all these features simultaneously. The proposed device achieves, to the best of our knowledge, the widest operational bandwidth reported to date for a nanophotonic multimode and dual-polarization operation silicon power splitter. Our design relies on a symmetric Y-junction with a SWG-metamaterial-assisted transition. This novel structure incorporates two strategically designed SWG sections that leverage distinct geometrical features. An initial region based on widening SWG segments to ensure smooth modal transitions. The subsequent region features SWG segments that bifurcate and linearly reduce in width, enabling precise beam shaping through the output arms. This approach enables flexible control over the modal transition at the junction tip, ensuring that the modes propagating through each branch do not experience an abrupt refractive index discontinuity. By judiciously designing the metamaterial effective refractive index, both energy reflection and radiation can be reduced compared to conventional counterparts. Three-dimensional simulations show that the proposed SWG-assisted Y-junction yields minimal excess loss (< 0.2 dB) for the transverse-electric fundamental mode ($TE_0$) and the first-order mode ($TE_1$) over a remarkable 700 nm bandwidth (1300 – 2000 nm), and less than 0.3 dB for the fundamental transverse-magnetic mode ($TM_0$) within the 1300 – 1800 nm wavelength range. Experimental measurements validate the simulation results, confirming low excess losses for the three modes in the 1430 – 1680 nm wavelength range (limited by our characterization setup) along with robust fabrication tolerances for etching errors up to ±10 nm.

## 2. Principle of operation and device design

Symmetric Y-junctions, comprising a stem waveguide that branches into two diverging arm waveguides of equal width, rely on adiabatic mode evolution. An ideal adiabatic transition in such devices (i.e. with small branching angles) results in minimal losses and polarization- and wavelength-agnostic spectral response for the supported modes [50]. Nevertheless, each spatial mode exhibits some specificities in their operation. Let us consider a symmetric Y-junction that supports both fundamental and first-order modes in the stem waveguide. Due to the inherent symmetry of the device, the power distribution of the fundamental mode is evenly split between the two output arms in phase. Conversely, for the first-order mode, power is also equally divided between the fundamental mode of each output arm, but the relative phase between them is π. A more detailed description of modal evolution in Y-junctions can be found in Reference [51], by considering the stem as a symmetric three-layer slab waveguide, and the arms a symmetric five-layer slab waveguide.

The resulting losses depend on the operating mode and can be ascribed to two main mechanisms. The first mechanism relates to the wavefront inclination as the mode in the stem waveguide propagates through the junction, resulting from gradual tilting in the propagation direction. The second mechanism involves changes in the mode profile along the branching. Thus, losses for the first-order mode are typically lower than for the fundamental mode. The field profile of the fundamental mode has one lobe in the center of the stem waveguide that transforms into a two-lobe distribution (with one lobe in each arm waveguide), while the field profile of the first-order mode has two lobes with a zero-field value at the center of the stem waveguide [52]. Notice that the first loss mechanism is mainly determined by tilt angle (and hence arm length), whereas the second is driven by the minimum separation between the arms, which in turn is limited by the resolution of fabrication processes. This contributes to further increase radiation loss and introduces additional back reflections, particularly affecting the fundamental mode as its field profile is maximum at the center of the stem waveguide and aligns with the abrupt tip at the junction.

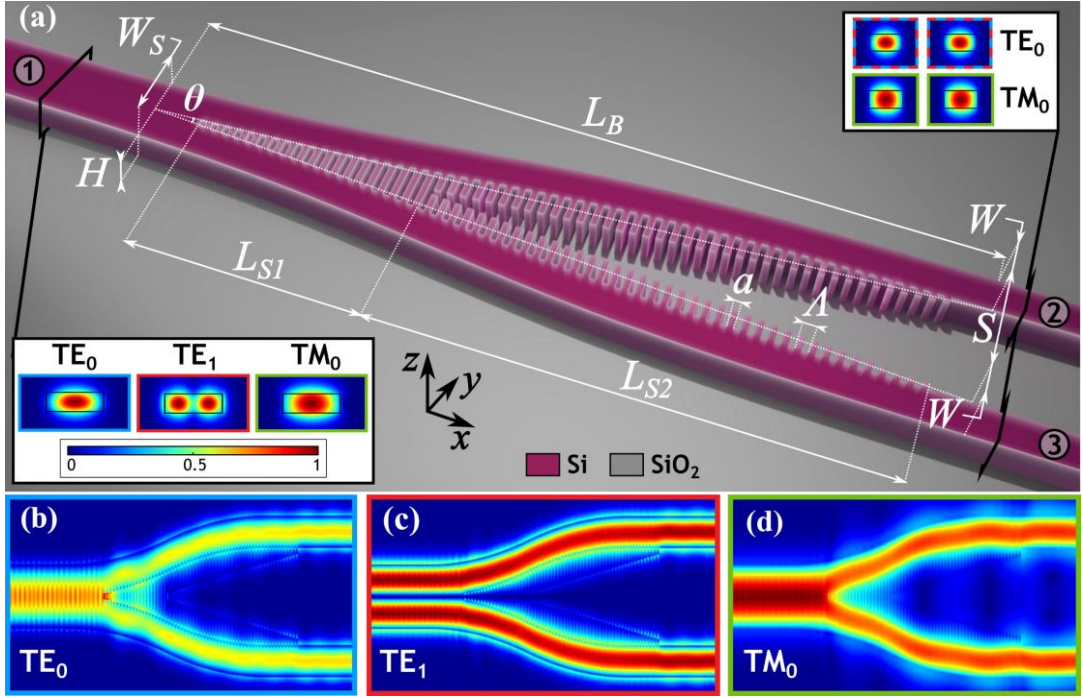

**Fig. 1.** (a) Three-dimensional schematic of the SWG-assisted Y-junction comprising two S-shaped arms and an SWG-metamaterial-assisted transition. The SiO$_2$ upper cladding is not shown for clarity. Insets: field distribution ($|E_y|$) of the fundamental and first-order transverse-electric modes in the stem and arm waveguides. 3D FDTD simulations of the field ($|E_y|$) propagation through the device at a wavelength of 1550 nm for (b) TE$_0$ (c) TE$_1$ and (d) TM$_0$ modes. The normalized colorbar applies to all field distributions.

To overcome these limitations and mitigate tip-induced losses of symmetric Y-junctions, we engineer the stem-to-arms transition by introducing an SWG metamaterial between the splitter arms, as depicted in Fig. 1(a). Although other approaches using SWG metamaterials in Y-junctions have been demonstrated [30,46], none of them achieve dual polarization operation and, when operating for higher-order modes, their losses or bandwidth is compromised. The proposed design includes two SWG sections, each precisely engineered to exploit different geometries to effectively mitigate losses for multimode and dual polarization operation over an ultra-broad bandwidth. First, an initial region comprising SWG segments that gradually widen. The succeeding region is distinguished by SWG segments that bifurcate and linearly reduce in width, enabling precise beam shaping through the output arms and ensuring a gradual transformation and separation of the beam profile with minimal losses. The incorporation of SWG metamaterials directly enhances the optical transition from the stem to the arms by leveraging their additional degrees of freedom to control mode confinement. This approach mitigates the abrupt refractive index discontinuity between silicon (Si) and the silicon dioxide (SiO$_2$) upper cladding at the junction, ensuring a smooth modal evolution. The subwavelength metamaterials implemented between the arms ensure a nearly invariant propagation constants near the junction, which is essential for minimizing losses due to mode transformation from a single-lobe to a dual-lobe profile in fundamental modes for both polarizations. The SWG period ($\Lambda$) must be selected to be small enough to prevent Bragg reflections, and large enough to comply with MFS constraints. The proposed Y-junction also incorporates arms with an S-bend shape in order to reduce the branching angle ($\theta$) at the junction, hence minimizing losses due to wavefront tilt mismatch.

The device was designed and optimized for a silicon-on-insulator platform with a 220-nm-thick silicon core layer through rigorous three-dimensional finite-difference time-domain (3D FDTD) simulations. An MFS of 100 nm was considered in the entire structure to ensure compatibility with state-of-the-art deep-ultraviolet (DUV) fabrication processes and higher resolution processes such as electron-beam lithography

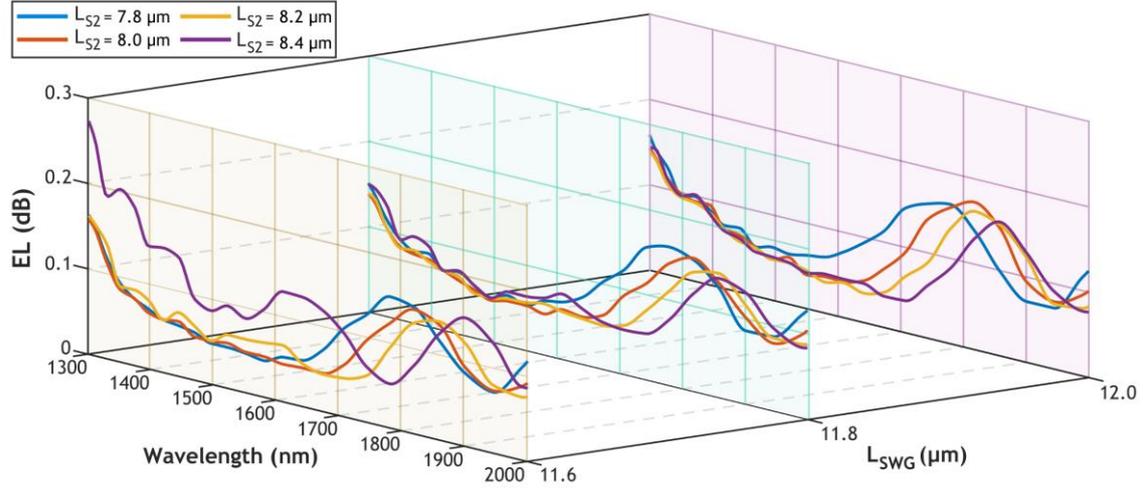

**Fig. 2.** Simulated excess loss for the TE$_0$ mode as a function of the wavelength calculated through 3D FDTD simulations for different values of $L_{SWG}$ and $L_{S2}$.

[53]. The splitter consists of a stem waveguide of width $W_S = 1$ µm that supports the fundamental and first-order transverse-electric (TE) modes and the fundamental transverse-magnetic (TM) mode, and two S-bend output arms with SWG metamaterials between them. The width of the arm waveguides ($W$) was set to 0.5 µm to be compatible with typical interconnection waveguides. The stem waveguide width further restricts second-order mode propagation, hence preventing possible coupling of odd-to-odd higher-order modes. The final arm separation ($S$) is set to 1.5 µm to ensure compactness and prevent power coupling between the output waveguides. An adiabatic transition [50] was achieved by setting the branching angle $\theta \approx 7°$, resulting in an arm length $L_B = 12.3$ µm.

The SWG period was set to 200 nm to prevent Bragg and radiation regimes, with a duty cycle (DC = $a/\Lambda$) of 50% to maximize the MFS. The SWG section of length $L_{SWG}$ is divided into two sections to take better advantage of the grating geometry. The first section of length $L_{S1}$ comprises SWG segments located at the beginning of the branching, whose width increases with arm spacing. In the second section of length $L_{S2}$, the SWG segments are split along the waveguide of each output arm, while linearly decreasing in size from $W_1$ to $W_2$. This allows for improved light splitting by using SWG metamaterials to manipulate the beam according to the shape and spacing of the arm waveguides. The effect of this topology on the evolution of the different modes can be seen in Figs. 1(b) to 1(d) [51].

S-parameters were calculated through Lumerical 3D FDTD simulations and used to evaluate the device's excess loss: EL = $10 \cdot \log_{10}(|S_{21}|^2 + |S_{31}|^2)$, where $|S_{21}|$ and $|S_{31}|$ are the transmission coefficients from the stem to each arm of the Y-junction. The TE$_0$ mode excess loss was obtained for different values of $L_{SWG} = L_{S1} + L_{S2}$ and $L_{S2}$, as shown in Fig. 2. Since TE$_1$ shows greater resilience to junction tip variations, optimization was focused on TE$_0$ performance. Trade-offs between minimal loss values and curve flatness across the widest operating wavelength range can also be observed. A larger $L_{SWG}$ tends to shift the best performing region to shorter wavelengths, which can be compensated for by increasing $L_{S2}$. Based on these results, the length of the two SWG regions were judiciously selected as $L_{SWG} = 11.8$ µm and $L_{S2} = 8.2$ µm in order to yield minimal losses

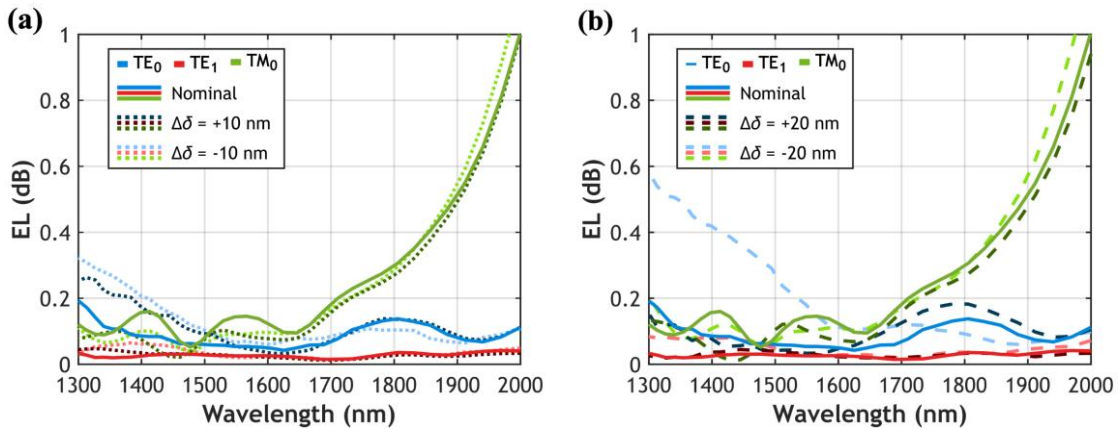

**Fig. 3.** Simulated excess loss as a function of the wavelength calculated through 3D FDTD simulations for TE$_0$ (blue), TE$_1$ (red) and TM$_0$ (green) modes for nominal (solid curves) and biased designs with (a) $\Delta\delta = \pm 10$ nm (dotted lines) and (b) $\Delta\delta = \pm 20$ nm (dashed lines).

across the entire simulated wavelength range (i.e., 1300 – 2000 nm).

Fig. 3 shows the simulated EL of the optimized SWG-assisted Y-junction design. To provide precise simulation results, mesh dimensions were set to $dx = 20$ nm, $dy = 12.5$ nm, $dz = 22$ nm and the time step was set to 0.0317 fs. The $TE_0$ mode excess loss ($EL_{TE0}$) and the $TE_1$ mode excess loss ($EL_{TE1}$) remain consistently below 0.2 dB and 0.07 dB, respectively, both over an ultra-broad bandwidth of 700 nm (from 1300 to 2000 nm). Therefore, the operation of the device for these modes fully covers the E, S, C, L and U telecom bands, partially the O band, and extends up to 2000 nm. Moreover, the device also offers polarization-independent spectral response at wavelengths below 1800 nm, but the operational bandwidth for $TM_0$ mode is limited at longer wavelengths due to leakage losses to the substrate. Excess loss for the $TM_0$ mode ($EL_{TM0}$) is lower than 0.3 dB within the wavelength range from 1300 to 1800 nm.

We also assessed the tolerances to fabrication deviations of the SWG-assisted Y-junction to etching errors of $\Delta\delta = \pm 10$ nm and $\Delta\delta = \pm 20$ nm from the nominal design, while operating for the three modes: $TE_0$, $TE_1$, and $TM_0$, as depicted in Fig. 3 (a) and (b), respectively. To account for potential under-etching and over-etching during the fabrication process, adjustments were made to the device geometry by considering these errors as absolute variations in the waveguide dimensions. Specifically, the length of the SWG silicon segments ($a$) was modified to $a' = a + \Delta\delta$ and the widths $W_1$ and $W_2$ became $W_1' = W_1 + \Delta\delta$ and $W_2' = W_2 + \Delta\delta$. Note that modifying the length and width of the silicon segments accordingly varies the grating duty cycle. The widths of the stem and arm waveguides were also adjusted to $W_S' = W_S + \Delta\delta$ and $W' = W + \Delta\delta$, respectively. Simulations demonstrate robust fabrication tolerances for the proposed splitter, without performance degradation across the entire 700 nm bandwidth for all $TE_0$, $TE_1$, and $TM_0$ modes for $\Delta\delta = \pm 10$ nm. For $\Delta\delta = \pm 20$ nm, only $TE_0$ performance at lower wavelengths is impacted, but even in this worst case scenario, EL remains below 0.6 dB in the 700 nm bandwidth and below 0.3 dB in a 500 nm bandwidth.

## 3. Fabrication and experimental characterization

The device was fabricated within a multiproject wafer run in a commercial foundry. A silicon-on-insulator wafer with a 220-nm Si layer and a 2-µm $SiO_2$ buried oxide (BOX) layer was used. The integrated circuit pattern was defined using a 100 keV electron-beam lithography system and then etched into the silicon layer using an anisotropic reactive ion etching process. To protect the silicon devices, a 2.2-µm-thick $SiO_2$ cladding was deposited by chemical vapor deposition. The facets of the chip were deep etched, providing smooth trenches around the perimeter of the chip. To achieve efficient fiber-chip edge coupling, high-performance and broadband SWG edge couplers that operate for both TE and TM polarizations were included in the input and output facets of the chip [35].

For the experimental characterization, we employed two tunable lasers to sweep the 1420 – 1505 nm and the 1505 – 1680 nm wavelength ranges, respectively. The lasers were connected to an optical stage to control the polarization state of incident light precisely. The polarization control stage comprised a three-paddle fiber polarization controller, in addition to an air-fiber bench polarization controller that included a linear polarizer and a half-wave plate. Subsequently, incident light was coupled into the chip using a polarization-maintaining lensed optical fiber. The light exiting the chip was collimated with a 40× microscope objective, and the resulting beam was directed towards a Glan-laser calcite polarizer. Finally, the beam was focused on a germanium photodetector for data acquisition and subsequent processing. The experimental setup is illustrated schematically in Fig. 4.

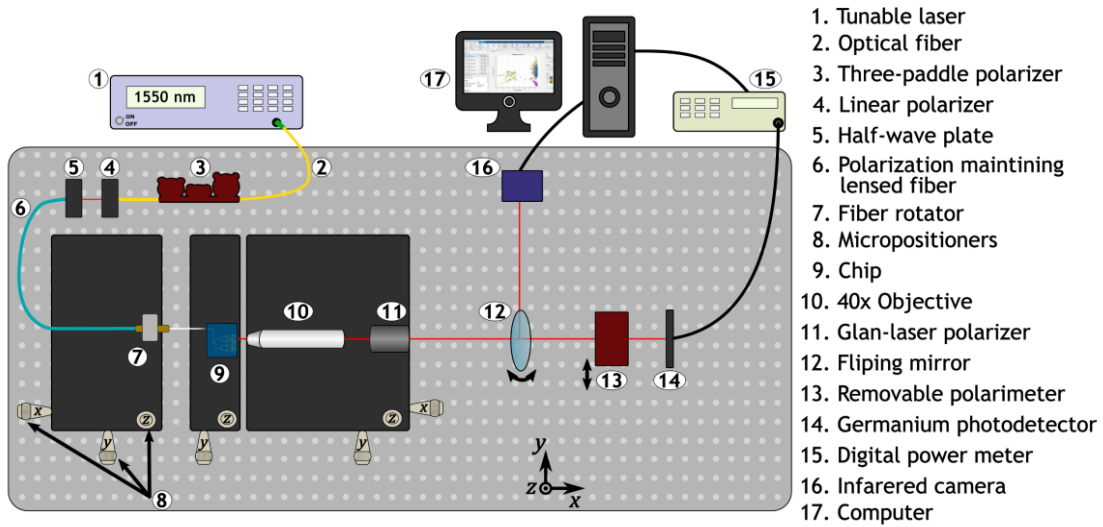

**Fig. 4.** Schematic of the experimental setup used for device characterization.

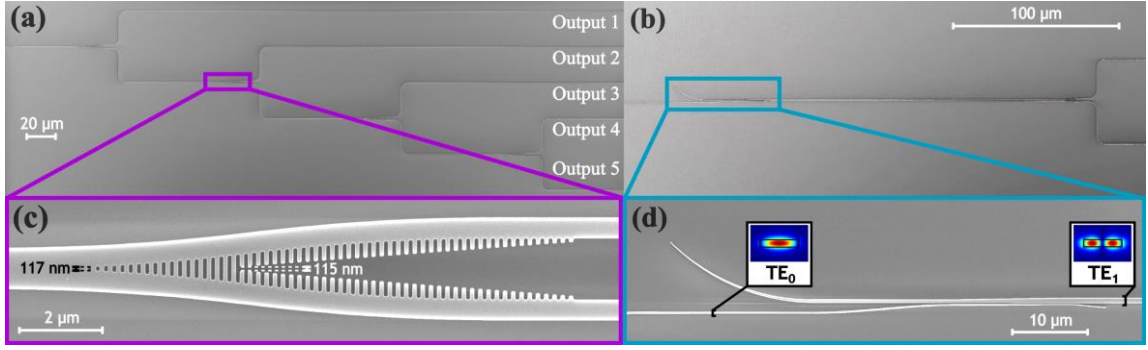

**Fig. 5.** SEM images of (a) the cascaded structure employed for $TE_0$ characterization and (b) the cascaded structure including mode multiplexers for $TE_1$ characterization. Insets (c) and (d) show detailed views of the SWG-assisted Y-junction and the auxiliary mode multiplexer (with $|E_y|$ field distribution of the fundamental and first-order transverse-electric modes), respectively.

*3.1 Measurements of transverse-electric modes*

The performance of the SWG-assisted splitters for $TE_0$ mode was assessed using cascaded structures composed of five splitters. In each splitter, the upper arm was routed to the chip output, while the lower arm was connected to the next splitter. This arrangement resulted in a structure with one input and five outputs, as illustrated in the scanning electron microscope (SEM) image in Fig. 5(a), with an inset of the fabricated SWG-assisted Y-junction in Fig. 5(c). To evaluate the operation of the proposed splitter for the $TE_1$ mode, the same cascaded structures used for the $TE_0$ mode characterization were incorporated, with the addition of a mode multiplexer preceding each of the five splitters [54]. Fig. 5(b) shows an SEM image of the mode multiplexer and the SWG-assisted Y-junction, with an inset of the mode multiplexer in Fig. 5(d). In this structure, when light was injected into the lower branch of the mode multiplexer, $TE_1$ mode was excited in the stem waveguide of the SWG-assisted Y-junction. This $TE_1$ mode was then split into the $TE_0$ modes of the two arms, with the upper arm directed to the chip output and the lower arm directed to the following mode multiplexer. Reference structures with four mode multiplexers in back-to-back configuration were also included to independently determine their losses, as well as reference structures to determine losses from input/output coupling and wire waveguide propagation.

The normalized transmission of the five outputs operating for $TE_0$ and $TE_1$ is presented in Fig. 6(a), (b), respectively, over the entire measured wavelength range spanning from 1430 to 1680 nm. The power of each output is in very good agreement with its corresponding target value, i.e., output 1 is centered at 0.5, output 2 at 0.25, output 3 at 0.125, and outputs 4 and 5 at 0.0625. Note that the small discontinuity observed at the wavelength of 1505 nm is due to the change of laser to perform the wavelength scan.

Excess losses for both $TE_0$ and $TE_1$ were accurately obtained by linear regression, fitting the power values measured from outputs 1 to 5. The resulting $EL_{TE0}$ and $EL_{TE1}$ are shown in Fig. 7(a), (b), respectively, with solid lines. Excess losses remain below 0.2 dB over the entire measured bandwidth from 1430 to 1680 nm for both $TE_0$ and $TE_1$ modes. Experimental results validate simulation predictions with a remarkable consistency (see Fig. 3). We also calculated the power uniformity (PU) at the first output, defined as the difference between the maximum and minimum output power within the measured bandwidth: $PU = 10 \cdot \log_{10}(P_{1,max}/P_{1,min})$. The resulting PU for $TE_0$ operation is 0.88 dB, while for $TE_1$, it is further reduced to 0.66 dB.

To validate the robustness of the proposed splitter against fabrication deviations, biased devices with induced dimensional variations of $\Delta\delta = \pm 10$ nm with respect to the nominal design were also fabricated. These devices were similarly arranged in cascaded structures for $TE_0$ and $TE_1$ operation. The measured ELs of biased devices are shown in Fig. 7(a) for $TE_0$ and in Fig. 7(b) for $TE_1$, indicated by dotted lines. In both cases, the largest performance degradation occurs for over-etching errors ($\Delta\delta = -10$ nm). Still, losses are lower than 0.4 dB for both modes in the 250 nm measured bandwidth. Note that slight negative loss values may appear due to the device's low losses, as measured transmittance can exceed that of reference structures due to minor variations in coupling efficiency or waveguide imperfections. These results demonstrate the resilience of the proposed SWG-engineered Y-junction to consistently operate with minimal losses, even in the presence of fabrication-related imperfections.

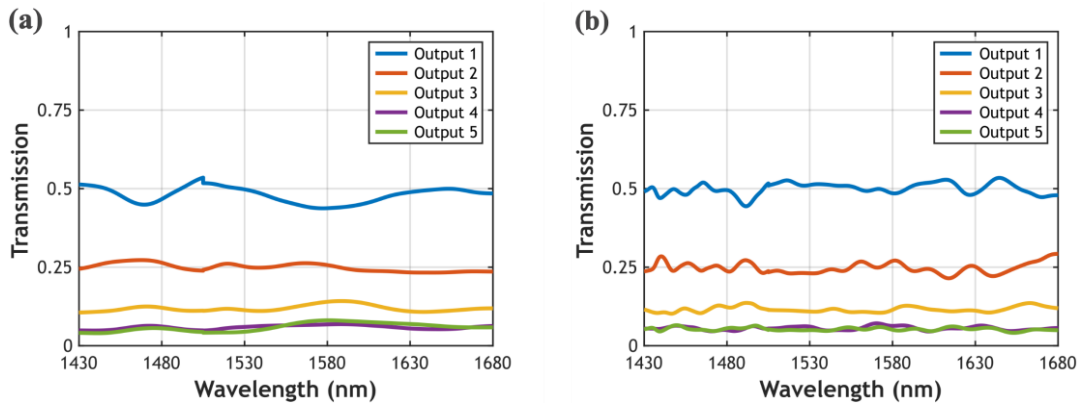

**Fig. 6.** Normalized transmission spectrum of outputs 1 to 5 measured using cascaded SWG-assisted Y-junctions operating for (a) $TE_0$ and (b) $TE_1$.

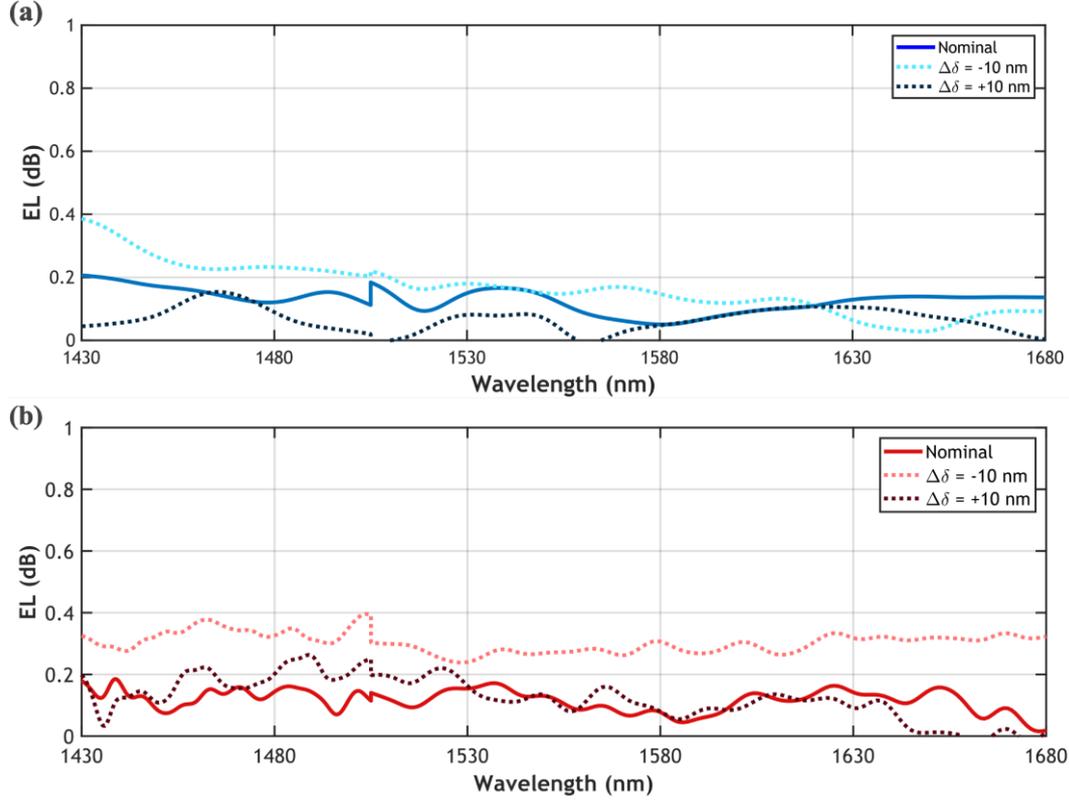

**Fig. 7.** Measured excess losses as a function of the wavelength for the nominal design (solid lines) and for devices with deviations of $\Delta\delta = \pm 10$ nm (dotted lines) for (a) $TE_0$ and (b) $TE_1$ operation. Note that the discontinuity at 1505 nm results from the laser change for the wavelength scan.

*3.2 Measurement of the fundamental transverse-magnetic mode*

The operation of the SWG-assisted Y-junction for $TM_0$ mode operation was also examined. The same cascaded structures used for $TE_0$ characterization were employed, but TM polarization was excited at the chip input. Fig. 8 shows $EL_{TM0}$, obtained by the same linear regression analysis. $EL_{TM0}$ measurements for the nominal device are represented with a solid line, while biased devices with dimensional variations of $\Delta\delta = \pm 10$ nm are represented with dotted lines. The upper limit of the measured bandwidth has been limited because the test structures were originally designed for TE polarization with

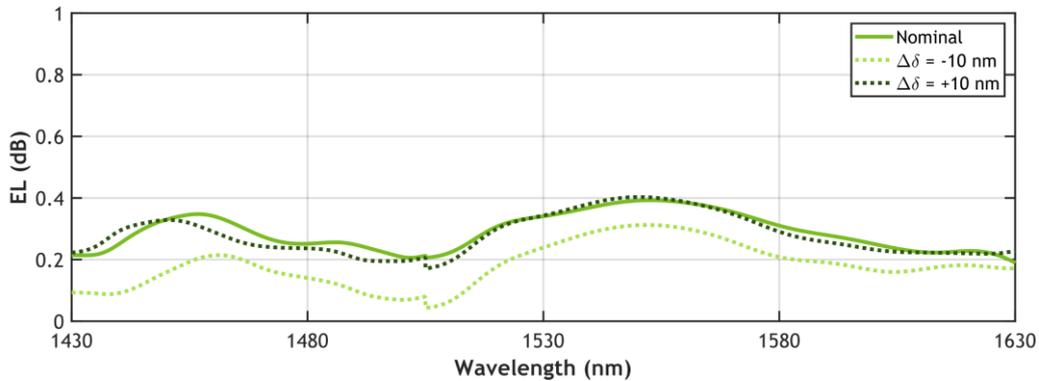

**Fig. 8.** Measured excess losses as a function of the wavelength for the nominal design (solid lines) and for devices with deviations of $\Delta\delta = \pm 10$ nm (dotted lines) for $TM_0$ operation. Note that the discontinuity at 1505 nm results from the change of laser source.

Table 1. Comparison of state-of-the-art multimode power splitters using 220-nm-thick silicon core.

| Device | Length (µm) | MFS (nm) | Modes | $EL_{sim}$ (dB) | $BW_{sim}$ (nm) | $EL_{exp}$ (dB) | $BW_{exp}$ (nm) |
|---|---|---|---|---|---|---|---|
| MMI [39] | 48 | 120 | $TE_0$ | < 0.40 | 100 (1500 – 1600) | < 0.65 | 100 (1500 – 1600) |
|  |  |  | $TE_1$ |  |  |  |  |
| Y–junction [46] | 29.15 | 40 | $TE_0$ | < 0.10 | 70 (1530 – 1600) | < 0.50 | 70 (1530 – 1600) |
|  |  |  | $TE_1$ † |  |  |  |  |
| Directional Coupler [55] | 2.62 | 100 | $TE_0$ | < 1.00 | 80 (1520 – 1600) | < 1.00 | 50 (1540 – 1590) |
|  |  |  | $TM_0$ |  |  |  |  |
| Adiabatic tapers [20] | 20 | 30 | $TE_0$ | < 0.05* | 500 (1200 – 1700) | < 0.19 | 70 (1530 – 1600) |
|  |  |  | $TM_0$ | < 0.10* | 425 (1200 – 1625) | < 0.14 |  |
| Y–junction [29] | 5 | 100 | $TE_0$ | < 0.45 | 790 (1250 – 2040) | NF | NF |
|  |  |  | $TM_0$ | < 1.00 | 385 (1200 – 1585) |  |  |
| Pixelated meta-structure [48] | 5.4 | 120 | $TE_0$ | < 0.83 | 445 (1588 – 2033) | NF | NF |
|  |  |  | $TE_1$ |  |  |  |  |
| Tapered coupler [56] | 13 | 150 | $TE_0$ | < 0.07 | 100 (1500 – 1600) | < 1.50 | 80 (1520 – 1600) |
|  |  |  | $TE_1$ † | < 0.41 |  |  |  |
| MMI [44] | 35.9 | 100 | $TE_0$ | < 1.00 | 160 (1260 – 1420) | NF | NF |
|  |  |  | $TM_0$ |  |  |  |  |
| Y–junction [23] | 14 | 120 | $TE_0$ | < 0.10 | 150 (1500 – 1650) | < 0.25 | 100 (1500 – 1600) |
|  |  |  | $TM_0$ |  |  | < 0.23 |  |
| Y–junction [31] | 41.4 | 100 | $TE_0$ | < 0.30 | 350 (1350 – 1700) | < 0.50 | 260 (1420 – 1680) |
|  |  |  | $TE_1$ | < 0.45 | 300 (1300 – 1600) | < 1.50 | 170 (1420 – 1590) |
| Y–junction (This work) | 12.3 | 100 | $TE_0$ | < 0.20 | 700 (1300 – 2000) | < 0.20 | 250 (1430 – 1680) |
|  |  |  | $TE_1$ | < 0.20 |  |  |  |
|  |  |  | $TM_0$ | < 0.30 | 500 (1300 – 1800) | < 0.40 | 200 (1430 – 1630) |

a) Values marked with an asterisk correspond to values estimated from Figures. b) Devices annotated with a dagger symbol also operate for higher-order modes. NF, Not Fabricated; BW, Bandwidth; EL, Excess Loss; MFS, Minimum Feature Size.

5-µm-radii bends, resulting in radiation for wavelengths beyond 1630 nm when operating with TM polarization. Nevertheless, the device exhibits an excellent performance for $TM_0$ mode splitting, with $EL_{TM0}$ below 0.4 dB over a wide bandwidth from 1430 to 1630 nm. The excess losses for the splitters with modified dimensions at $\Delta\delta = +10$ nm are very similar to the losses of the nominal device, achieving an $EL_{TM0} < 0.4$ dB over the same wavelength range. On the other hand, when the variations are $\Delta\delta = -10$ nm, the maximum losses in the spectrum are reduced to 0.3 dB. These results demonstrate the polarization-agnostic response of the proposed splitter, as well as the resilience to typical etching deviations for TM polarization operation.

## 4. Discussion and conclusions

In conclusion, we have proposed and experimentally demonstrated a metamaterial-assisted silicon 3-dB power splitter for multimode operation over a record broad bandwidth. The proposed design exploits the unique properties of subwavelength grating metamaterials to smooth the abrupt transition at the tip of conventional Y-junctions while flexibly controlling the confinement of the propagating modes. This approach enables efficient power splitting from the stem waveguide to the S-shaped arms with an outstanding performance for $TE_0$, $TE_1$ and $TM_0$ mode operation. 3D FDTD simulations predict an excess loss lower than 0.2 dB for the $TE_0$ mode and below 0.07 dB for the $TE_1$ mode over the wavelength range of 700 nm, which extends from 1300 to 2000 nm. The device also operates for the fundamental TM mode with simulated excess loss under 0.3 dB in the 1300 – 1800 nm wavelength range.

Measurements validate the splitting performance of the device with excess losses below 0.2 dB for both fundamental and first-order TE modes in the 1430 to 1680 nm wavelength range, limited by the lasers available in our setup. Measurements for the $TM_0$ mode also show low excess loss, with less than 0.4 dB over a 200 nm bandwidth from 1430 to 1630 nm. At the operational wavelength of 1550 nm, the proposed SWG splitter exhibits a measured excess loss of 0.15 dB for $TE_0$ mode, 0.13 dB for $TE_1$ mode, and 0.39 dB for $TM_0$ mode. These discrete values also compares favorably against the state of the art. Only the design presented in [20] achieves slightly lower excess losses (<0.11 dB for $TE_0$ and <0.08 dB for $TM_0$), at the expense of a narrower operational bandwidth of just 70 nm. Another significant advantage of the proposed SWG-assisted Y-junction is its robustness against fabrication imperfections. The proposed splitter reliably maintains its high performance even when subjected to fabrication-related geometric variations, which was validated by both simulations and experiments. SEM images of the fabricated devices across the chip showed minimal deviations from the design, with only minor corner rounding observed. Despite these slight variations, the experimental results aligned closely with simulation predictions, exhibiting deviations of less than 0.1 dB across the measured range for all three optical modes, thereby validating the robustness and reproducibility of the design. This provides a good indicator of the quality of the simulation in order to predict behavior beyond our measurable range.

For the sake of comparison, in Table 1 we review and compare the reported device with several state-of-the-art multimode power splitters based on 220-nm-thick silicon core. To ensure a fair comparison, it should be noted that different authors define the operational bandwidth according to different loss thresholds, indicated therein. Among stand-out devices, Ozcan et al. recently reported a polarization-independent Y-junction, achieving low losses (EL < 0.25 dB) and a device length of only 14 µm, but the bandwidth is limited to 100 nm [23]. The widest experimental bandwidth (170 nm) is achieved by [31], still 80 nm smaller than our $TE_1$ performance, and still at the expense of greater losses, larger footprint, and no TM operation. The widest simulated performance is achieved by [29], although no experimental validation is provided, no $TE_1$ operation is provided, and bandwidth is defined for a larger loss threshold. Compared to these alternatives, the SWG-assisted splitter in this work represents a substantial advance over current state-of-the-art devices and outperforms them in terms of optical bandwidth with at least 100 nm improvement compared to most prior experimental demonstrations, without increasing the length and maintaining ultra-low losses. Moreover, these results are achieved while preserving an MFS of 100 nm, compatible with state-of-the-art DUV fabrication processes.

The SWG-assisted symmetric Y-junction presented in this work is expected to unlock new possibilities for several applications requiring both broad bandwidth and precise power control, especially for multimode applications and/or those with multiple or cascaded power splitting stages. These include both traditional and emerging areas of silicon photonics, including high-capacity data transmission across multiple optical bands, light detection and ranging, programmable photonics, multi-target spectroscopy, quantum key distribution, and optical sensing, among others. Moreover, the proposed design strategy can be applied to other high-index contrast material platforms. In terms of scalability, the design could potentially support higher-order modes by increasing the width of the stem and arms waveguides, though this would require a challenging re-optimization of the geometric parameters to account for potential mode coupling between even-to-even and odd-to-odd modes [50, 51]. Higher-order TM modes would also benefit from a thicker core layer to prevent leakage into the substrate. We believe that the low losses, ultra-broad bandwidth, and multimode operation along with fabrication tolerances of the proposed 3-dB splitter will also be essential for the implementation of advanced and complex systems requiring a high precision in the handling of optical signals.

**CRediT authorship contribution statement**

**Raquel Fernández de Cabo:** Conceptualization, Software, Formal Analysis, Methodology, Investigation, Validation, Visualization, Data Curation, Writing – original draft. **David González-Andrade:** Conceptualization, Data Curation, Formal Analysis, Methodology, Supervision, Writing – review & editing. **Pavel Cheben:** Conceptualization, Writing – review & editing. **Aitor V. Velasco:** Conceptualization, Data Curation, Methodology, Resources, Funding Acquisition, Supervision, Project Administration, Writing – review & editing.

**Declaration of Competing Interest**

The authors declare that they have no known competing financial interests or personal relationships that could have appeared to influence the work reported in this paper.

**Acknowledgements**

This work has been funded by Spanish Ministry of Science and Innovation (PID2020-115353RA-I00); the

Spanish State Research Agency and the European Social Fund Plus under grant PRE2021-096954, and the European Union's Horizon Europe (HADEA grant agreement No. 101135523).